\documentclass[conference]{IEEEtran}

\usepackage{cite}
\usepackage{amsmath,amsthm,amssymb,amsfonts}
\usepackage{graphicx}
\usepackage{textcomp}
\usepackage{xcolor}
\usepackage{color}
\usepackage{setspace}

\usepackage{mathrsfs}
\usepackage{pifont}
\usepackage{bm}
\usepackage{pgfplots}
\usepackage{tikz}
\usetikzlibrary{arrows}
\usepackage{subfigure}
\usepackage{graphicx,booktabs,multirow}
\usepackage{upgreek}
\usepackage{bm}
\usepackage{mathrsfs}
\usepackage{algorithm}
\usepackage{algorithmicx}
\usepackage{algpseudocode}
\newcommand{\vct}[1]{{\bm #1}}

\newcommand{\cmark}{\text{\ding{51}}}
\newcommand{\xmark}{\text{\ding{55}}}

\definecolor{colorhkust}{RGB}{20,43,140}
\definecolor{colortsinghua}{RGB}{116,52,129}
\definecolor{color1}{RGB}{128,0,0}

\newtheorem{theorem}{Theorem}

\newtheorem{proposition}{Proposition}

\mathchardef\re="023C
\mathchardef\im="023D

\IEEEoverridecommandlockouts
\begin{document}

\title{A Graph Neural Network Approach for Scalable Wireless Power Control}

%

\author{
   \IEEEauthorblockN{Yifei Shen$^\dagger$, Yuanming Shi$^\star$, Jun Zhang$^\ddagger$, and Khaled B. Letaief$^{\dagger}$}\\
  \IEEEauthorblockA{$^\dagger$Dept. of ECE, The Hong Kong University of Science and Technology, Hong Kong\\
  	$^\star$School of Information Science and Technology, ShanghaiTech University, Shanghai 201210, China\\
  	$^\ddagger$ Dept. of EIE, The Hong Kong Polytechnic University, Hong Kong\\
  	Email:{ yshenaw@connect.ust.hk, shiym@shanghaitech.edu.cn, jun-eie.zhang@polyu.edu.hk, eekhaled@ust.hk} 
  	\thanks{This work was supported by the Hong Kong Research Grants Council under Grant No. 16210719.}}

}

\maketitle

\begin{abstract}
Deep neural networks have recently emerged as a disruptive technology to solve NP-hard wireless resource allocation problems in a real-time manner. However, the adopted neural network structures, e.g., multi-layer perceptron (MLP) and convolutional neural network (CNN), are inherited from deep learning for image processing tasks, and thus are not tailored to problems in wireless networks. In particular, the performance of these methods deteriorates dramatically when the wireless network size becomes large. In this paper, we propose to utilize graph neural networks (GNNs) to develop scalable methods for solving the power control problem in $K$-user interference channels. Specifically, a $K$-user interference channel is first modeled as a complete graph, where the quantitative information of wireless channels is incorporated as the features of the graph. We then propose an interference graph convolutional neural network (IGCNet) to learn the optimal power control in an unsupervised manner. It is shown that one-layer IGCNet is a universal approximator to continuous set functions, which well matches the permutation invariance property of interference channels and it is robust to imperfect channel state information (CSI). Extensive simulations will show that the proposed IGCNet outperforms existing methods and achieves significant speedup over the classic algorithm for power control, namely, WMMSE.
\end{abstract}

\begin{IEEEkeywords}
Resource allocation, geometric deep learning, graph neural networks, wireless networks.
\end{IEEEkeywords}
\begin{table*}[t]
	
	\selectfont  
	\centering
	
	\caption{A comparison of different methods for power control in the $K$-user interference channel.} 
	
	\resizebox{1\textwidth}{!}{
		\begin{tabular}{|c|c|c|c|c|c|c|}  
			\hline  
			&  MLP \cite{sun2018learning} & DPC \cite{lee2018deep} & PCNet \cite{liang2018towards}  & Spatial Convolution \cite{cui2018spatial} & Graph Embedding \cite{lee2019graph}  & This paper \cr\hline
			Neural networks used&MLP&CNN&MLP&Spatial Convolution \cite{cui2018spatial}&Structure2Vec \cite{dai2016discriminative}&Proposed IGCNet\cr\hline
			Training scheme&Supervised&Supervised or Unsupervised&Unsupervised&Supervised or Unsupervised&Supervised or Unsupervised&Unsupervised\cr\hline 
			Scalability&$\xmark$&$\xmark$&$\xmark$&$\cmark$&$\cmark$&$\cmark$\cr\hline
			Time complexity &$\mathcal{O}(K^2)$&$\mathcal{O}(K^2)$&$\mathcal{O}(K^2)$&$\mathcal{O}(K+N^2)$& $\mathcal{O}(K^2)$ &$\mathcal{O}(K^2)$\cr\hline
			Sample complexity&Medium&Medium&Medium&Large&Small&Small\cr\hline
			Ability to incorporate instantaneous CSI&$\cmark$ &$\cmark$&$\cmark$&$\xmark$&$\xmark$&$\cmark$\cr\hline 
			Ability to solve weighted problems&$\cmark$ &$\xmark$&$\cmark$&$\xmark$&$\xmark$ & $\cmark$\cr\hline

	\end{tabular}}
	\label{tab:methods}
\end{table*}
\vspace{-0.5em}
\section{Introduction}\label{sec:intro}
Effective resource allocation plays a crucial role for performance optimization in wireless networks. However, typical resource allocation problems, such as power control \cite{chiang2008power,shi2014group}, are non-convex and computationally challenging. Moreover, they need to be solved in a real-time manner to accommodate the time variation of wireless channels. Great efforts have been put to develop effective algorithms for wireless resource allocation, and design solutions have been obtained with powerful convex optimization based approaches. Nevertheless, the resulting algorithms still fall short, given the increasing density of wireless networks and the more stringent latency requirement of emerging mobile applications.

Inspired by the recent successes of deep learning, researchers have attempted to apply deep learning based methods to solve NP-hard optimization problems in wireless networks \cite{sun2018learning,lee2018deep,liang2018towards,shen2018lora,zappone2018model,lee2019learning}. As a classic wireless resource allocation problem, power control in the $K$-user interference channel has attracted most of the attention \cite{sun2018learning,lee2018deep,liang2018towards,cui2018spatial,lee2019graph}. The first attempts came from \cite{sun2018learning,lee2018deep}, which applied MLP and CNN, respectively, to approximate the classic weighted minimum mean square error (WMMSE) algorithm \cite{Shi2011An} and accelerate the computation. Unsupervised learning and an ensembling mechanism were employed in \cite{liang2018towards} to achieve better performance than the sub-optimal WMMSE algorithm. 

However, MLP and CNN, which are designed for image processing, may not be suitable for problems in wireless communication. In particular, the performance of these methods degrades dramatically when the network size becomes large. This is because MLP and CNN fail to exploit the underlying topology of wireless networks. To enable more efficient learning, spatial convolution \cite{cui2018spatial} and graph embedding \cite{lee2019graph} have been proposed to exploit the Euclidean geometry of the users' geolocations. These methods are scalable to large-size networks. However, they have major disadvantage, namely, they can not utilize the instantaneous channel state information (CSI), which can not be embedded into the Euclidean space. This leads to poor performance in fading channels. Another drawback is that they have trouble in dealing with problems with heterogeneity, e.g., weighted sum rate maximization. A comparison of the existing works for $K$-user interference channel power control is shown in Table \ref{tab:methods}.

Graph neural networks (GNNs) can effectively exploit non-Euclidean data \cite{wu2019comprehensive}, e.g., CSI. In this paper, to overcome the limitations mentioned above, we propose to employ GNNs for wireless power control in $K$-user interference channels. Specifically, a $K$-user interference channel can be naturally modeled as a complete graph, where the quantitative information of wireless networks, e.g., CSI, is incorporated as the features of the graph. Based on the principle of graph neural networks, we propose \emph{interference graph convolutional networks} (IGCNet) to learn the optimal resource allocation in an unsupervised manner. It is shown that IGCNet is a universal approximator of continuous set functions, which well preserves the permutation invariance property of the interference links. Extensive simulations will demonstrate that the proposed IGCNet not only outperforms the state-of-the art optimization-based WMMSE algorithm and existing learning-based methods under various system configurations, but also achieves significant speedup over WMMSE. Furthermore, we will show that the proposed IGCNet can handle estimation uncertainty, e.g., CSI uncertainty, both theoretically and empirically. For reproducibility, the code to produce the results in this paper has been made available on github\footnote{https://github.com/yshenaw/Globecom2019}.

\section{Preliminaries} 
\begin{figure*}[htbp]
	\centering
	\subfigure[An illustration of power control via an MLP. The channel matrix is first flattened into a 1D vector, which results in the structure information loss.]{
		\begin{minipage}[t]{0.47\linewidth}
			\centering
			\includegraphics[width=1\linewidth]{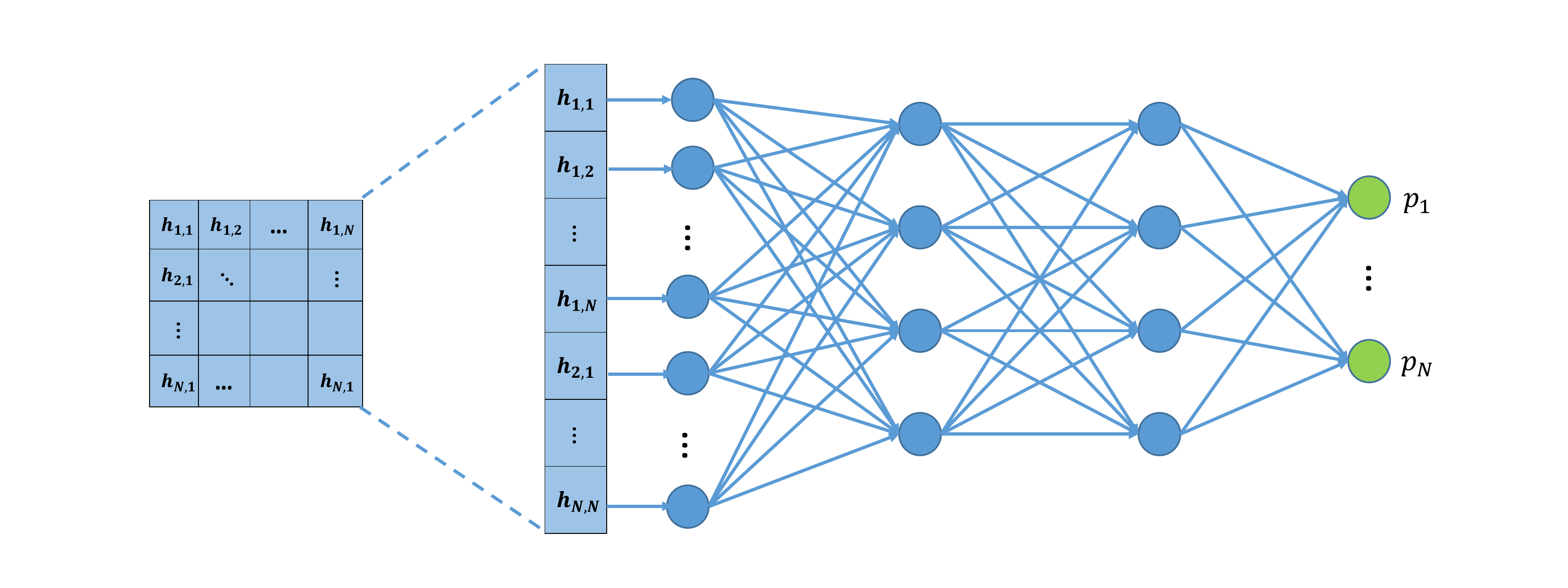}
			\label{fig:MLP} 
		\end{minipage}%
	}%
	\hspace{.1in}
	\subfigure[An illustration of power control via a CNN. Only nearby elements are put together in the 2D convolution.]{
		\begin{minipage}[t]{0.47\linewidth}
			\centering
			\includegraphics[width=1\linewidth]{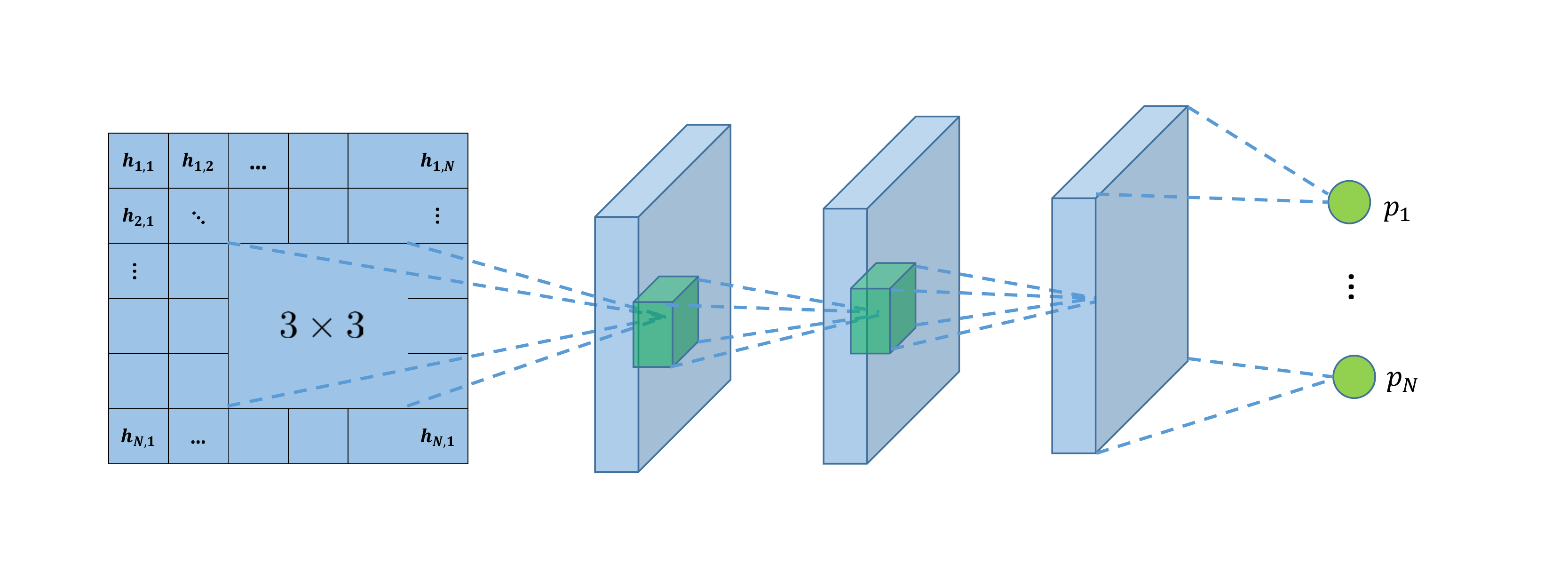}
			\label{fig:CNN}
		\end{minipage}%
	}%
	\centering
	\label{fig:PC}
	\caption{CNN and MLP for $K$-user interference channel power control.}
\end{figure*}

\subsection{System Model} \label{sec:sys_model}
We consider the power control problem in a $K$-user interference channel with $K$ single-antenna transceiver pairs. The received signal at the $k$-th receiver is given by 
$$y_k = h_{kk}s_k + \sum_{j \neq k} h_{kj}s_j + n_k,$$
where $h_{kk} \in \mathbb{C}$ denotes the direct-link channel between the $k$-th transmitter and receiver, $h_{kj} \in \mathbb{C}$ denotes the cross-link channel between transmitter $j$ and receiver $k$, $s_k \in \mathbb{C}$ denotes the data symbol for the $k$-th receiver, and $n_k \sim \mathcal{CN}(0,\sigma_k^2)$ is the additive Gaussian noise.

The signal-to-interference-plus-noise ratio (SINR) for the $k$-th receiver is given by
$$\text{SINR}_k = \frac{|h_{kk}|^2p_k}{\sum_{i\neq k}|h_{ki}|^2p_i+\sigma_k^2},$$
where $p_k=\mathbb{E}[s_k^2]$ is the power of the $k$-th transmitter, and $0 \leq p_k \leq P_{\text{max} }$.

Denote $\vct{p}=[p_1,\cdots,p_K]$ as the power allocation vector. The objective is to find the optimal power allocation to maximize the weighted sum rate, and the problem is formulated as
\begin{equation}\label{eq:sys_mod}
\begin{aligned}
&\underset{\vct{p}}{\text{maximize}}
& & \sum_{k=1}^{K} w_k \log_2 \left(1+ \text{SINR}_k \right) \\
& \text{subject to}
& & 0 \leq p_k \leq P_{\text{max} }, \forall k,
\end{aligned}
\end{equation}
where $w_k$ is the weight for the $k$-th pair. The channel matrix is defined as $\bm{H} = [\bm{h}_1,\cdots,\bm{h}_K]^T$ and $\bm{h}_i = [h_{1i},\cdots,h_{Ki}]^T, i=1,\cdots,K$.

This problem is known to be NP-hard \cite{chiang2008power}. Although several optimization-based methods have been proposed in \cite{Shi2011An,shen2018fractional}, they are computationally demanding, and thus cannot be applied for real-time implementation \cite{sun2018learning}. To alleviate the computation burden while achieving near-optimal performance, machine learning based methods have been proposed. Specifically, MLP \cite{sun2018learning,liang2018towards} and CNN \cite{lee2018deep} have been used to approximate the input-output mapping of this problem. The optimization-based methods involve many iterations, with each iteration having a time complexity of $\mathcal{O}(K^2)$. In contrast, the total complexities of these learning-based methods are $\mathcal{O}(K^2)$, and thus they can achieve significant speedups.

\subsection{Existing Approaches' Limitations}
In this subsection, we identify the performance deterioration phenomenon of existing methods using MLP or CNN \cite{sun2018learning,lee2018deep,liang2018towards}.

Fig. \ref{fig:PC} illustrates MLP and CNN based approaches for power control. From the numerical experiments in \cite{sun2018learning,liang2018towards}, we observe a performance loss when $K$ gets larger. For example, in \cite{sun2018learning}, the performance gap to the WMMSE algorithm is $3\%$ when $K=10$ and becomes $12\%$ when $K=30$. From the perspective of approximation theory, an MLP with a sufficient number of parameters can learn anything if we have sufficient training samples \cite{hornik1989multilayer}. However, in practice, there are lots of redundancy in an MLP because it is fully connected. Such redundancy causes overfitting and makes it difficult to train. This is the reason for the performance loss when the input and output dimensions are large.

CNN has demonstrated its effectiveness in solving such performance deterioration problems in image analysis applications \cite{brendel2019approximating}, but it is not effective for wireless power control. Specifically, for images, the geometric property means that adjacent pixels are meaningful to be considered together \cite{brendel2019approximating}. In CNN, a 2D convolution kernel is applied to each patch (adjacent pixels) in the image. The weights in the neural network are shared among different patches. This leads to a significant reduction in the number of parameters, which leads to a lower sample complexity and also makes it easy for training. This accounts for the superior performance of CNN in image processing. Unfortunately, the geometric property in images does not hold for a channel matrix since a $3 \times 3$ patch does not contain any specific meaning for the power control problem. Thus, although using CNN can reduce the number of parameters, it suffers from a large performance degradation, as will be further shown in Section \ref{sec:Gaussian}.

There have been some attempts to leverage the geometry of users' geolocation to achieve scalability, i.e., the ability to deal with large-size wireless systems. One study \cite{cui2018spatial} applied the idea of convolution in the spatial domain. Specifically, the whole considered area is first divided into $N$-by-$N$ grids, followed by computing the number of active users in each grid as its density. The spatial convolution is a convolution operator on the density grid. In this scenario, the neighbor grids are useful because the nearest users will cause the strongest interferences. One major drawback of this work is that it requires a large number of samples for training. To address this issue, \cite{lee2019graph} proposed to use distance quantization and graph embedding. However, spatial convolution and distance quantization are merely operating on the distances and can not incorporate instantaneous CSI. Thus, it results in poor performance when fading exists (as shown in Table V in \cite{cui2018spatial}). Furthermore, they are not able to deal with the weighted problems.

In the next section, we will discuss the geometric properties of the $K$-user interference channels and design the corresponding neural network.

\section{Learning Optimal Power Control on Interference Graph}
In this section, we first model a $K$-user interference channel as a complete graph, followed by a brief introduction to GNNs. Under the framework of GNNs, we propose IGCNet to learn the optimal power control on an interference graph in an unsupervised manner. The theoretical analysis for IGCNet is presented at the end of this section.

\subsection{Graph Representation and Geometric Properties} \label{sec:graph}
\begin{figure}[htb]
	\centering
	\includegraphics[width=0.48\textwidth]{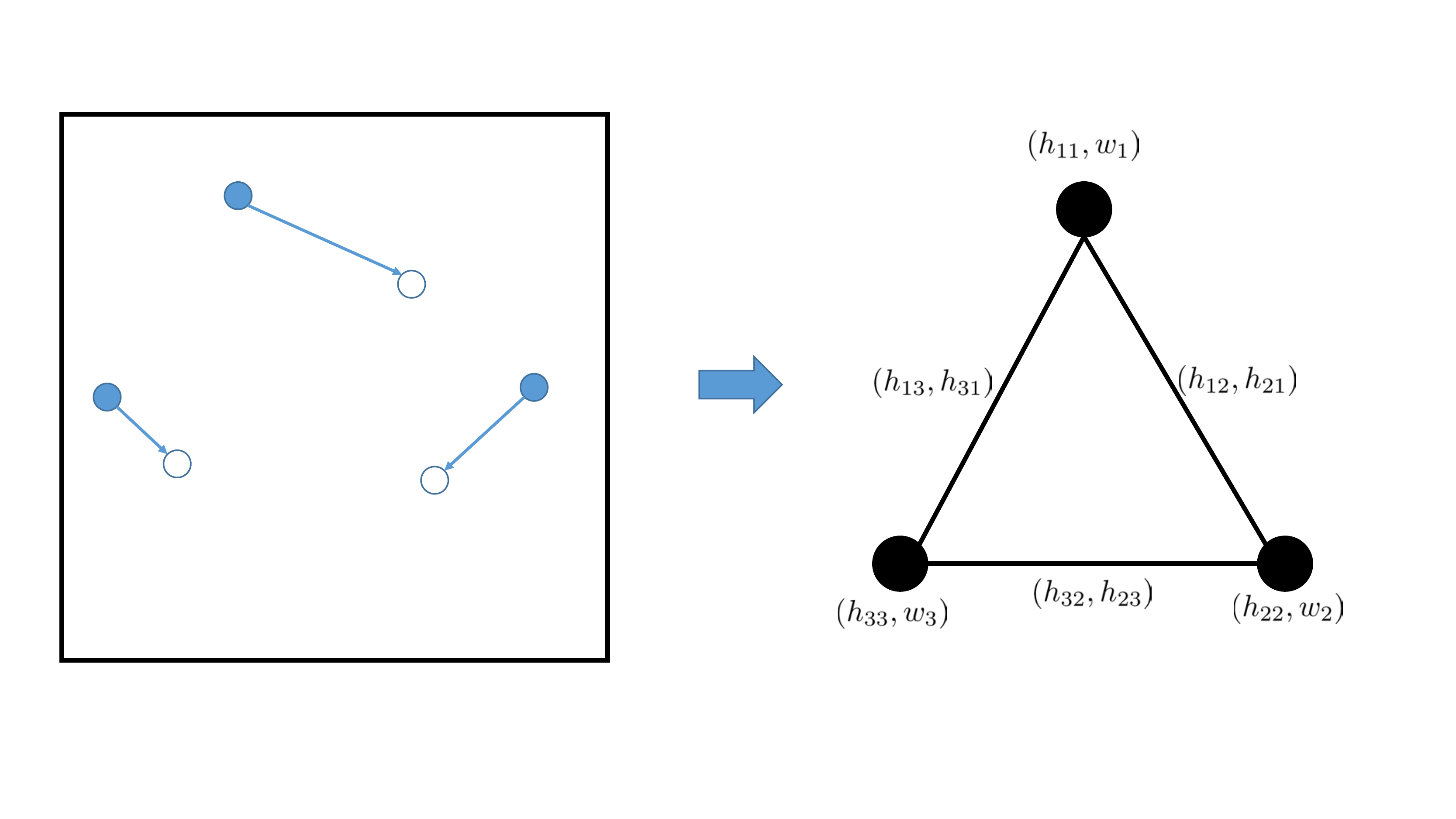}
	\caption{The $3$-user interference channel and the corresponding graph.}
	\label{fig:MDG}
\end{figure}

In this subsection, we model the $K$-user interference channel as a complete graph with vertex and edge labels. We view the $i$-th transmitter-receiver pair as the $i$-th vertex. The vertex label contains the state of the direct channel and the weight of the $i$-th pair, i.e., $(h_{ii},w_i)$. One edge between two vertices indicates an interference link, with label as the states of the interference channels $h_{ij}$ and $h_{ji}$. An illustration of a 3-user interference channel is shown in Fig. \ref{fig:MDG}. 

We next discuss the geometry of the interference channel by looking at the map from the channel matrix and weights to the optimal power control vector.

\begin{proposition}\label{prop:pi}
	For a given $i$, let $f_i(\cdot,\cdot)$ denote the function that maps the channel matrix and the weights to the optimal power allocation of the $i$-th transmitter, i.e., $p_i^* = f_i(\bm{H},\bm{w})$, and let $\bm{\Pi}$ denote any permutation matrix satisfying $(\bm{\Pi}^T \bm{H} \bm{\Pi})_{ii} = h_{ii}$. Then,  $p_i^* = f_i(\bm{H},\bm{w}) = f_i(\bm{\Pi}^T \bm{H} \bm{\Pi},\bm{\Pi}^T\bm{w})$.
\end{proposition}

This can be interpreted as the unordered property of interference channels : It is the collection of interference channel coefficients instead of the ordering of these coefficients that matter. The irrelevance in the ordering leads to the permutation invariance property of the channel matrix. This property suggests that only considering the neighborhood elements, such as in CNN, is meaningless because the elements are no longer close to each other after the permutation. This invariance property indicates that all the edges with the same end node are homogeneous, and will allow us to share weights among all the edges of a node. In other words, we can restrict the hypothesis space of the designed neural network for one node to the space of set functions, which leads to GNNs.

\subsection{Graph Neural Networks}
In this subsection, we give a brief introduction to GNNs, and one can refer to \cite{xu2018powerful,wu2019comprehensive} for a more detailed information. GNNs deal with learning problems with graph data or non-Euclidean data. There are many sucessful applications of GNNs such as recommendation systems \cite{ying2018graph} and solving combinatorial problems \cite{li2018combinatorial}. GNNs utilize the graph structure of data, the node features, and the edge features to learn a good representation of the vertices. Like MLP or CNN, GNNs have layer-wise structures. In each layer, for each vertex, GNNs update the representation of this node by aggregating features from its edges and its neighbor vetices. Specifically, the update rule of the $k$-th layer at vertex $v$ in GNNs is 
\begin{equation}\label{eq:gnn}
\small
\begin{aligned}
&\bm{\alpha}^{(k)}_v = \text{AGGREGATE}^{(k)}(\{\bm{\beta}^{(k-1)}_u: u \in \mathcal{N}(v)  \}, \{\bm{\gamma}_x: x \in \mathcal{E}(v)\} ) \\
&\bm{\beta}^{(k)}_v = \text{COMBINE}^{(k)}(\bm{\beta}^{(k-1)}_v, \bm{\alpha}^{(k)}_v)
\end{aligned}
\end{equation}
where $\mathcal{N}(v)$ denotes the set of the neighbors of $v$, $\mathcal{E}(v)$ denotes the set of edges with $v$ as one end node, $\text{AGGREGATE}(\cdot)$ and $\text{COMBINE}(\cdot)$ are two functions, $\beta^{(k)}_v$ denotes the $k$-th layer's output feature of vertex $v$, and $\alpha^{(k)}_v$ is an intermediate variable.

The design of the two functions in GNNs is crucial and leads to different kinds of GNNs \cite{xu2018powerful}. The most popular GNNs are listed as follows.
\begin{enumerate}
	\item \textbf{Graph Convolutional Network} \cite{kipf2016semi}: It uses the mean pooling and relu as the aggregation and combination functions,
	$$\bm{\beta}^{(k)}_v = \text{RELU} (\bm{W}^{(k)} \cdot \frac{1}{|\mathcal{N}(v)|+1}\sum_{u}\bm{\beta}^{(k-1)}_u)$$
	where $u \in \mathcal{N}(v) \cup \{v\}$ and $\{\bm{W}^{(k)} \}$ are the weight matrices to be learned and $\text{RELU}(x)=\text{MAX}(0,x)$.
	\item \textbf{Structure2Vec} \cite{dai2016discriminative}: It uses the sum pooling and relu as the aggregation and combination functions,
	$$\bm{\beta}^{(k)}_v = \text{RELU} \left( \bm{W}_1 \bm{\beta}^{(k-1)}_v + \bm{W}_2 \sum_{u} \bm{\beta}^{(k-1)}_u \right), $$
	where $u \in \mathcal{N}(v)$, and $\bm{W}_1$ and $\bm{W}_2$ are the weight matrices to be learned.
	\item \textbf{Graph Isomorphism Network} \cite{xu2018powerful}: It uses the MLP and sum pooling as the aggregation and combination functions, 
	$$\bm{\beta}^{(k)}_v = \text{MLP}^{(k)} \left( (1+\epsilon^{(k)}) \bm{\beta}^{(k-1)}_v + \sum_{u} \bm{\beta}^{(k-1)}_u \right)$$
	where $u \in \mathcal{N}(v)$, $\text{MLP}^{(k)}$ is an MLP, and different MLPs are used in different layers.
\end{enumerate}

\subsection{Interference Graph Convolutional Networks}
\begin{figure*}[htb]
	\centering
	\includegraphics[width=0.8\textwidth]{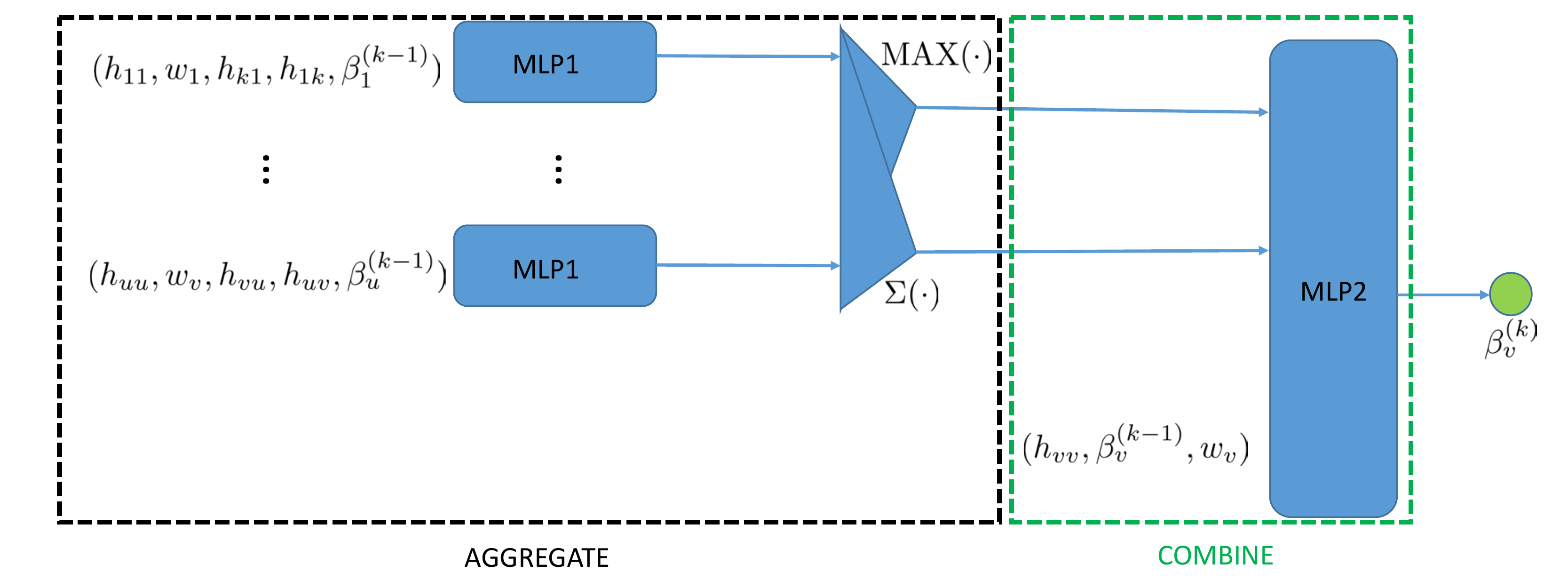}
	\caption{The structure of aggregation and combination functions in the proposed IGCNet.}
	\label{fig:ICCN}
\end{figure*}

In this paper, we design the aggregation and combination functions of a GNN following two principles based on the observations in Section \ref{sec:graph}. First, the designed neural network should capture the permutation invariance property of the interference channel, as stated in Proposition \ref{prop:pi}. Second, the designed neural network should be robust to the inaccurate measurements, e.g., imperfect CSI, which is critical for the practical implementation. Thus, the proposed aggregation functions (neural networks) should not only be a good approximator of set functions, but also robust to the corruptions of edge labels. For the first requirement, the idea is to use a symmetric function on transformed elements in the point set to approximate a general function defined on the set \cite{qi2017pointnet}:
$$\text{AGGREGATE}(\{\bm{x}_1,\cdots,\bm{x}_n\}) \approx g(o(\bm{x}_1),\cdots,o(\bm{x}_n)), $$
where $f:2^{\mathbb{R}^N} \rightarrow \mathbb{R}$, $o: \mathbb{R}^N \rightarrow \mathbb{R}^K$ and $g:\mathbb{R}^K \times \cdots \times \mathbb{R}^K \rightarrow \mathbb{R}$ is a symmetric function. For the implementation of this paper, we use a $3$-layer MLP as $o(\cdot)$, $\sum_{i}(\cdot)$ and $\text{max}(\cdot)$ as $g(\cdot)$, and a $3$-layer MLP as $\text{COMBINE}(\cdot)$. Specifically, the update rule is 
\begin{equation}
\begin{aligned}
	& \bm{\gamma}_{u,v}^{(k)} = \text{MLP1}(h_{uv},h_{vu},w_v,h_{uu}, \bm{\beta}^{(k-1)}_u) \\
	& \bm{\alpha}^{(k)}_v = \text{CONCAT}\left(\text{MAX}_u(\bm{\gamma}_{u,v} \}),\sum_{u} \bm{\gamma}_{u,v} \right), u \in \mathcal{N}(v) \\
	& \bm{\beta}^{(k)}_v = \text{MLP2}(\bm{\alpha}^{(k)}_v,h_{vv},\bm{\beta}^{(k-1)}_v,w_{v}),
\end{aligned}
\end{equation}
where $\text{MAX}(\{\cdot\})$ is to take the largest value in a set, $\text{MLP1}$ and $\text{MLP2}$ denote two different MLPs, $\text{CONCAT}$ denotes the operation that concatenates two vectors together, and $\gamma_{u,v}^{(k)}$ denotes the feature vector of the edge connecting vertex $u$ and vertex $v$. It will be shown in the next subsection that the proposed aggregation function also satisfies the second requirement. An illustration of the proposed network structure and parameter setting is shown in Fig. \ref{fig:ICCN}. 

The loss function adopted is the negative sum rate, as in \cite{liang2018towards,lee2018deep},
$$\ell = -\mathbb{E}_{\bm{H}} \left\{ \sum_{k=1}^{K} w_k \log_2 \left(1+ \frac{|h_{kk}|^2p_k(\theta)}{\sum_{i\neq k}|h_{ki}|^2p_i(\theta)+\sigma_k^2} \right) \right\},$$
where $\theta$ denotes the parameters of the neural networks and $p_i(\theta)$ is the power value generated by the neural networks. Note that this loss function is differentiable and can be directly optimized by stochastic gradient descent. Thus, IGCNet is an unsupervised method and it only needs the channel matrices as samples without labels for training.

The proposed IGCNet can also deal with other objective functions in the $K$-user interference channel. To achieve this goal, one can simply replace the loss function with the negative objective function to be maximized.

\subsection{Theoretical Analysis}
In this subsection, we show the universal approximation property and robustness of the proposed IGCNet. 

\begin{theorem} (Universal Approximation) \label{thm:uni}
	Suppose $f:\mathcal{X} \rightarrow \mathbb{R}$ is a continuous set function. Then, $\forall \epsilon > 0, \exists$ a continuous function h and a symmetric function $g(x_1,\cdots,x_n) = \zeta \circ \text{POOLING}$, such that $\forall S \in \mathcal{X}$,
	$$\left| f(S) - \zeta \left(\underset{x_i \in S}{\text{POOLING}}(\{h(x_i) \})  \right)\right| < \epsilon$$
	where $x_1,\cdots,x_n$ is the elements in $S$, $\zeta$ is a continuous function, and $\text{POOLING}$ is a pooling operation, i.e., $\text{POOLING}_{x_i \in S} (\{h(x_i) \}) = \text{MAX}_{x_i \in S}(\{h(x_i)\})$ or $\text{POOLING}_{x_i \in S}(\{h(x_i) \}) = \sum_{x_i \in S} h(x_i)$ or  $\text{POOLING}_{x_i \in S}(\{h(x_i) \}) = \frac{1}{|S|}\sum_{x_i \in S} h(x_i)$.
\end{theorem}

\begin{theorem} (Robustness)\cite{qi2017pointnet} \label{thm:robust}
	Suppose $\bm{u}:\mathcal{X} \rightarrow \mathbb{R}^p$ such that $\bm{u}=\text{MAX}_{x_i \in S }$ and $f=\gamma \circ \bm{u}$. Then, 
	
	(a) $\forall S, \exists \mathcal{C}_S, \mathcal{N}_S \in \mathcal{X}, f(T) = f(S) \text{ if } \mathcal{C}_S \subset T \subset \mathcal{N}_S$;
	
	(b) $|\mathcal{C}_S| \leq p$
\end{theorem}

Theorem \ref{thm:uni} states that one-layer IGCNet is a universal approximator of continuous set functions. Note that the aggregation function of GNNs is a set function. Thus, IGCNet has the aggregation function with the most powerful representation ability in the class of GNNs. Besides, it well respects the permutation invariance property of the interference channel because the set function is permutation invariant to the input. 

Theorem \ref{thm:robust} states that $f(S)$ remains the same up to the corruptions of the input if all the features in $\mathcal{C}_S$ are preserved and $\mathcal{C}_S$ only contains a bounded number of features, smaller than $p$. This means that only a small proportion of features are critical and IGCNet is robust to the corruptions of other features. The practical meaning is that IGCNet can provide a near-optimal solution even when some CSI is not available.

\section{Simulations} \label{sec:exp}

\subsection{Gaussian Interference Channel Power Control} \label{sec:Gaussian}
In this subsection, in order to demonstrate the effectiveness of IGCNet, we follow the system setting of \cite{sun2018learning,liang2018towards} to set up simulations. Under this system setting, all the weights are the same and the channel coefficients are taken from the standard normal distribution. In the experiment, three network setups $K \in \{10,20,30\}$ are considered. We mainly compare the proposed IGCNet with the following five benchmarks:
\begin{enumerate}
	\item WMMSE \cite{Shi2011An}: This is the most popular optimization-based algorithm for the $K$-user interference channel power control. It is also used as a benchmark in \cite{sun2018learning,lee2018deep,liang2018towards}.
	\item MLP \cite{sun2018learning}: It leverages MLP to learn the input-output mapping of WMMSE. 
	\item PCNet \cite{liang2018towards}: It employs MLP and an unsupervised loss function to learn near-optimal power allocation.
	\item DPC \cite{lee2018deep}: CNN and the unsupervised loss function are used in this method to learn a near-optimal power control.
	\item Baseline: We find a fixed proportion of pairs with the largest coefficients $w_i|h_{ii}|^2$, and set the power of these pairs as $P_{\text{max} }$, while the power for other pairs are set as $0$. This algorithm ignoring the interference is the simplest but effective heuristic algorithm and we shall use it as the baseline. 
\end{enumerate}

We do not compare with \cite{cui2018spatial} and \cite{lee2019graph} as they can not incorporate instantaneous CSI.

We generate $20000$ training samples, i.e., network realizations, to train MLP, PCNet, and DPC as in \cite{sun2018learning,lee2018deep} while the number of training samples used for IGCNet is $2000$. The test dataset contains $500$ network realizations. We use a $5$-layer IGCNet and adopt the adam optimizer with a learning rate of $10^{-3}$ to train IGCNet.  The simulation results are shown in Table \ref{tab:per_comp}. 

It is shown that the proposed IGCNet not only outperforms the learning-based method, but also achieves a better performance than WMMSE. We also see that the performance of IGCNet is stable while other learning-based methods suffer from performance degradation when $K$ increases. The difference between PCNet and IGCNet is that IGCNet utilizes the graph structure and the weights are shared. Thus, it suggests that leveraging the graph structure of the interference channel is useful for maintaining good performance when the network size is large. Besides, we observe that the performance gap between other learning-based methods and the baseline vanishes when $K=30$. This may imply that these models can hardly learn the impact of interference when $K$ is large. 

\begin{table}[htb]
	
	\selectfont  
	\centering
	
	\caption{Average sum rate under each setting. The results are normalized by the sum rate achieved by WMMSE.} 
	
	\resizebox{0.48\textwidth}{!}{
		\begin{tabular}{|c|c|c|c|c|c|c|}  
			\hline  
			 &  IGCNet & MLP &  PCNet & DPC & Baseline \cr\hline
			$K=10$ &$\bm{102.6\%}$&$98.2\%$&$101.4\%$&$95.1\%$&$89.1\%$\cr\hline
			$K=20$ &$\bm{102.7\%}$&$92.3\%$&$90.2\%$&$83.1\%$&$86.6\%$\cr\hline 
			$K=30$ &$\bm{102.4\%}$&$85.3\%$&$87.6\%$&$79.3\%$&$84.4\%$\cr\hline   
			
	\end{tabular}}
	\label{tab:per_comp}
\end{table}

We next test the performance of IGCNet in the weighted sum rate maximization. We take $w_i$ from the uniform distribution in $[0,1]$ and test the performance of IGCNet, WMMSE, and the baseline algorithm. The results are shown in Table \ref{tab:w_per_comp}. 

\begin{table}[htb]
	
	\selectfont  
	\centering
	
	\caption{Average sum rate under each setting. The results are normalized by the sum rate achieved by WMMSE.} 
	
	\resizebox{0.3\textwidth}{!}{
		\begin{tabular}{|c|c|c|c|c|c|c|}  
			\hline  
			&  IGCNet &   MLP&Baseline \cr\hline
			$K=10$ &$\bm{106.4\%}$&$93.7\%$&$92.5\%$\cr\hline
			$K=20$ &$\bm{106.9\%}$&$86.4\%$&$87.5\%$\cr\hline 
			$K=30$ &$\bm{104.7\%}$&$81.3\%$&$87.9\%$\cr\hline

	\end{tabular}}
	\label{tab:w_per_comp}
\end{table}

From Table \ref{tab:w_per_comp}, we see that the performance of IGCNet is better than other benchmarks. This demonstrates that IGCNet can handle the weighted problem without requiring a large amount of samples.

\subsection{Generalization}
An important test of the usefulness of neural network's design is its ability to generalize to different layouts and link distributions \cite{cui2018spatial}. In this subsection, we test the generalization performance of the proposed IGCNet in two different settings.

\subsubsection{Varying User Locations} \label{sec:diff} We first test the performance of the proposed algorithm under the situation where the locations of users in each sample vary. The transmitters are uniformly distributed in the square region $[0,100] \times [0,100]$ meters. The receivers are uniformly distributed within $[2,10]$ meters away from the transmitter. The adopted channel model is 
$$h_{ij} = 10^{-L(d_{ij})/20} \sqrt{\phi_{ij} s_{ij}} g_{ij}$$
where the path loss model is $L(d_{ij}) = 148.1+37.6\log_2(d_{ij} )$, $s_{kl}$ is the shadowing coefficient, the standard deviation of log-norm shadowing is $8$dB, $\phi_{ij}=9$dBi is the transmit antenna power gain, $g_{ij} \sim \mathcal{CN}(0,1)$ is the small scale fading, and the noise power is $\sigma_k^2 = -102$dBm. We use equal weights to test the performance of different algorithms, with the results shown in Table \ref{tab:w_per_geo}. We see that IGCNet also has a superior performance under the system settings with varying user locations.

\begin{table}[htb]
	
	\selectfont  
	\centering
	
	\caption{Average sum rate under each setting. The results are normalized by the sum rate achieved by WMMSE.} 
	
	\resizebox{0.4\textwidth}{!}{
		\begin{tabular}{|c|c|c|c|c|c|c|}  
			\hline  
			&  IGCNet &   MLP& PCNet&Baseline \cr\hline
			$K=10$ &$\bm{103.4\%}$&$83.7\%$&$86.4\%$&$75.9\%$\cr\hline
			$K=20$ &$\bm{104.4\%}$&$70.7\%$&$86.3\%$&$78.0\%$\cr\hline 
			$K=30$ &$\bm{104.2\%}$&$63.1\%$&$83.0\%$&$75.5\%$\cr\hline

	\end{tabular}}
	\label{tab:w_per_geo}
\end{table}

\subsubsection{Varying Distance Distribution} It was reported in \cite{cui2018spatial} that spatial convolution is sensitive to the link distance distribution. We also check the performance of IGCNet when the link distance distribution in the test is different from that in the training. We follow \cite{cui2018spatial} to set up the simulation. The link distance is uniformly distributed in $[2,10]$ meters during training. In the test, the link distance is uniformly distributed in $[l_r,u_r]$ meters, where $l_r$ is uniform in $[2,20]$ meters and $u_r$ is uniform in $[l_r,20]$ meters. The performance of IGCNet is $101.2\%$ compared to WMMSE in this situation. It shows that the performance of IGCNet under this setting is still good.

\subsection{Robustness}
In this subsection, the situation with partial CSI and noisy CSI are tested. We use the pre-trained model for $K=30$ in Section \ref{sec:diff}. 

\subsubsection{Partial CSI} 
\begin{figure}[htb]
	\centering
	\includegraphics[width=0.4\textwidth]{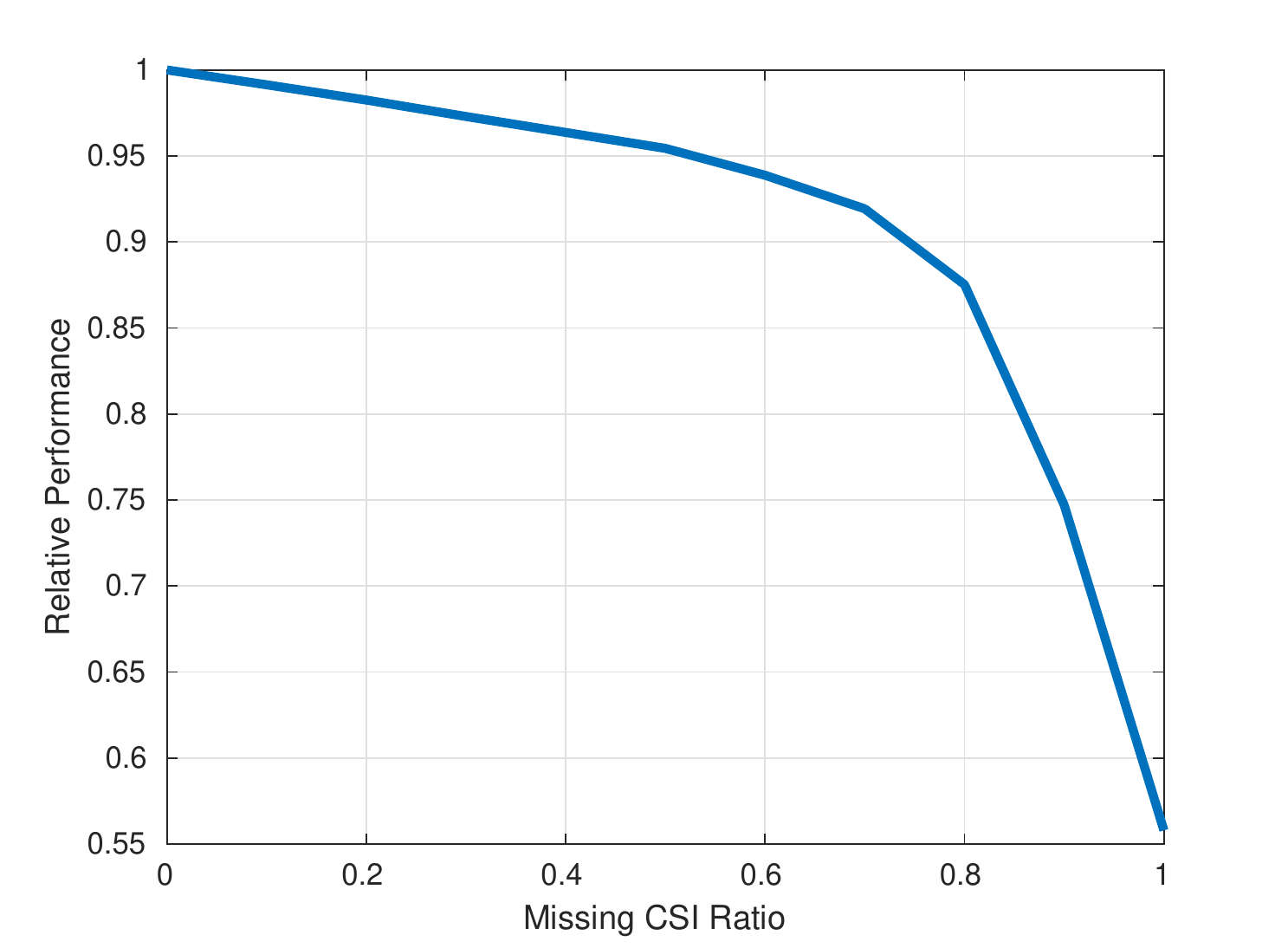}
	\caption{The relative performance versus the missing ratio of CSI.}
	\label{fig:partial}
\end{figure}

We simulate the situation where CSI of some links cannot be obtained. To test the performance under this partial CSI setting, we set a fixed proportion of $h_{ij}$ with the largest distance as $0$. We define the relative performance as the sum rate achieved by IGCNet with the partial CSI divided by the that achieved by the case with full CSI. The relative performance of IGCNet versus the missing CSI ratio is shown in Fig. \ref{fig:partial}. We see that IGCNet achieves $92\%$ performance of the full CSI case when $70\%$ of the links are set to $0$ in the available CSI, which verifies the robustness shown in Theorem \ref{thm:robust}.

\subsubsection{Noisy CSI} We simulate the situation where the CSI is inaccurate. We use the pre-trained model $K=30$ in Section \ref{sec:diff}. To test the performance under the noisy CSI setting, we use the noisy channel matrix $\hat{\bm{H}} = \bm{H} + \bm{N}$ as the input of IGCNet, where $\bm{N} \sim \mathcal{CN}(0,\sigma^2 I)$. We define the relative noise variance as $\eta = \frac{\sigma^2 \times K \times K}{\|\bm{H}\|_{F}^2}$. The relative performance of IGCNet versus the missing CSI ratio is shown in Fig. \ref{fig:partial}. We see that IGCNet achieves its $92\%$ performance when the relative noise variance is $10\%$. This suggests IGCNet is robust to CSI inaccuracy.

\begin{figure}[htb]
	\centering
	\includegraphics[width=0.4\textwidth]{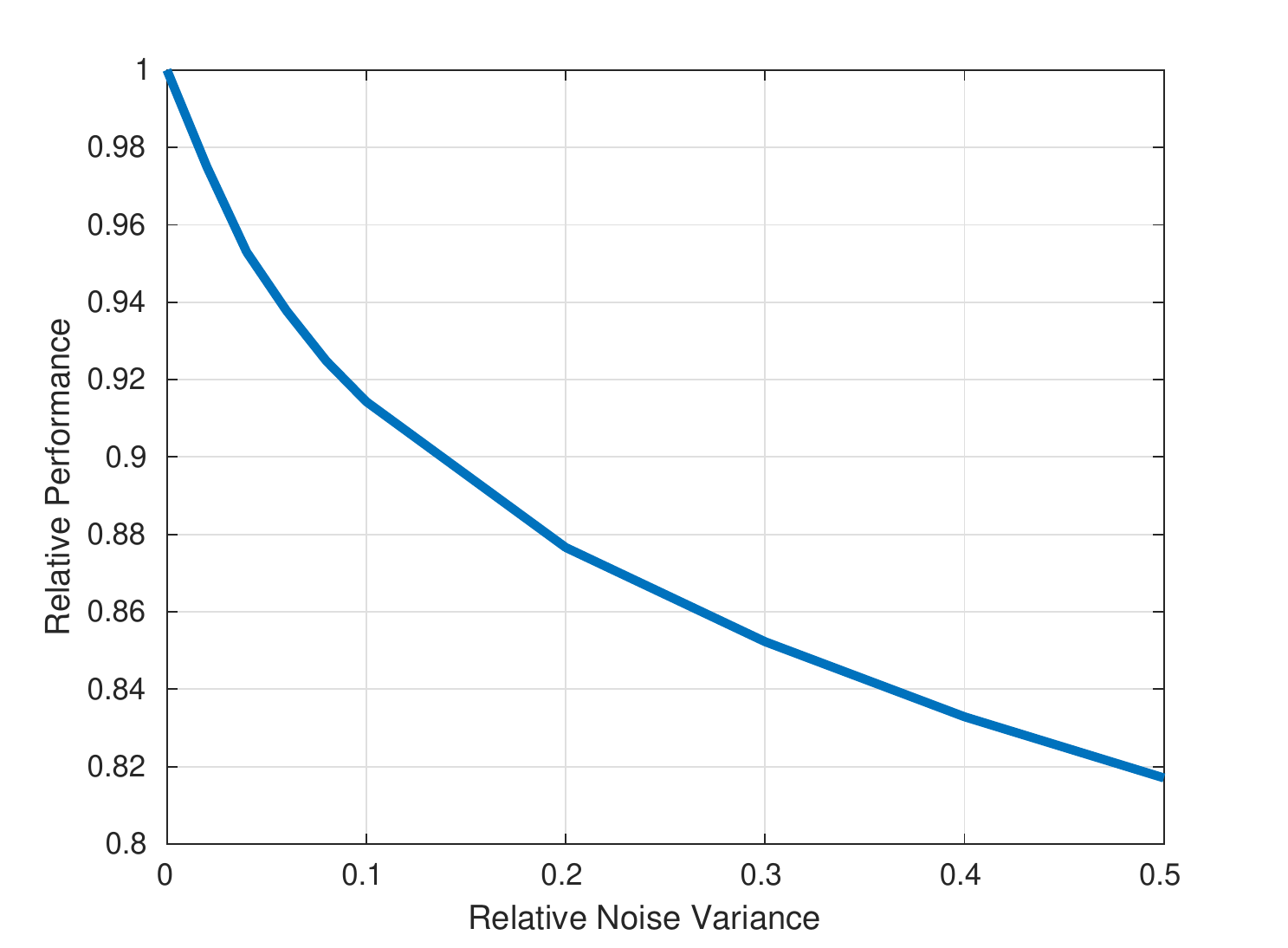}
	\caption{The relative performance versus the relative noise variance.}
	\label{fig:noisy}
\end{figure}


\subsection{Time Comparison}
It was reported in \cite{sun2018learning,liang2018towards} that learning-based methods have less computation time than the optimization-based methods. We also compare the average running time for WMMSE and IGCNet under the system setting in Section \ref{sec:diff}, as shown in Table \ref{tab:time_comp}. It can be concluded that IGCNet is significantly faster than WMMSE, up to about 65x speedup when $K=30$. This is because WMMSE involves many iterations, and each iteration has time complexity $O(K^2)$ while the total complexity of IGCNet is $O(K^2)$. 

\begin{table}[htb]
	
	\selectfont  
	\centering
	
	\caption{Average running time for the algorithms under each setting (in milliseconds).} 
	
	\resizebox{0.35\textwidth}{!}{
		\begin{tabular}{|c|c|c|c|c|c|c|}  
			\hline  
			&  $K=10$ &  $K=20$ & $K=30$ \cr\hline
			IGCNet &\textbf{0.14ms}&\textbf{0.27ms}&\textbf{0.48ms}\cr\hline
			WMMSE &9.31ms&24.1ms&31.4ms\cr\hline

	\end{tabular}}
	\label{tab:time_comp}
\end{table}

\subsection{Ablation Study}
We provide the ablation study of IGCNet in terms of the numbers of training samples and the number of layers of IGCNet. We assume Rayleigh fading and $K=30$.

We first study the impact of the numbers of training samples. We set the numbers of training samples as $\{50,200,500,2000,5000,20000\}$ and observe the test performance of IGCNet. The results are shown in Table \ref{tab:samples}. We see that the performance first increases then becomes stable when the number of samples increases. It also suggests $2000$ samples are sufficient for this setting, which are much less than the samples needed for previous works.

\begin{table}[htb]
	
	\selectfont  
	\centering
	
	\caption{Average sum rate of IGCNet with different numbers of training samples.} 
	
	\resizebox{0.48\textwidth}{!}{
		\begin{tabular}{|c|c|c|c|c|c|c|}  
			\hline  
			$\#$ Samples & $50$& $200$ &  $500$ & $2000$ & $5000$ & $20000$ \cr\hline
			Performance &$97.6\%$&$101.2\%$&$101.4\%$&$102.4\%$&$102.5\%$& $102.4\%$\cr\hline
			
	\end{tabular}}
	\label{tab:samples}
\end{table}

We then study the impact of the number of layers. $m$-hop information is gathered if a $m$-layer IGCNet is used. Intuitively, IGCNet with a larger number of layers will have better performance. We set the numbers of layers as $\{1,3,5,7,9\}$ and observe the test performance of IGCNet. 

\begin{table}[htb]
	
	\selectfont  
	\centering
	
	\caption{Average sum rate for IGCNet with different numbers of layers.} 
	
	\resizebox{0.48\textwidth}{!}{
		\begin{tabular}{|c|c|c|c|c|c|c|}  
			\hline  
			$\#$ Layers & $1$& $3$ &  $5$ & $7$ & $9$ \cr\hline
			Performance &$94.7\%$&$101.3\%$&$102.4\%$&$102.5\%$&$102.7\%$\cr\hline

	\end{tabular}}
	\label{tab:layers}
\end{table}

The performance improves as the number of layers increase. This is not surprising because IGCNet with more layers captures more information. We also see a huge performance gain from $1$-layer IGCNet to $3$-layer IGCNet, which shows that multi-hop information is crucial for the performance. From Table \ref{tab:per_comp} and Table \ref{tab:layers}, we find that $1$-layer IGCNet still outperforms MLP, PCNet, and DPC, which demonstrates the benefits of leveraging the graph structure.

In summary, the extensive simulations listed in Section \ref{sec:exp} have shown that 
\begin{enumerate}
	\item IGCNet not only outperforms other state-of-the-art learning-based methods, but also has better performance than the most popular optimization-based method, WMMSE, under various system configurations.
	\item IGCNet can generalize to different layouts and link distributions.
	\item IGCNet is robust to partial CSI and noisy CSI.
	\item IGCNet achieves significant speedups over WMMSE.
\end{enumerate}

\section{Conclusions}
In this paper, we developed a novel graph neural network for the $K$-user interference channel power control problem. The unique advantages include scalability, ability to incorporate instantaneous CSI, and ability to solve weighted problems. This is achieved by leveraging the geometric property and graph structure of the interference channels. For future directions, it will be interesting to test the effectiveness of IGCNet in other wireless resource allocation problems. We envisioned that machine learning based methods will play a critical role in future wireless networks \cite{letaief2019roadmap}.


\bibliographystyle{ieeetr}
\bibliography{ref}

\end{document}